\documentclass{bmvc2k}
\usepackage[normalem]{ulem}

\newcommand{\gh}[2]{\textcolor{olive}{\sout{#1}}{\textcolor{olive}{#2}}}

\usepackage{comment}
\usepackage{multirow}

\title{A cappella: Audio-visual Singing Voice Separation}

\addauthor{Juan F. Montesinos$^\ast$}{juanfelipe.montesinos@upf.edu}{1}
\addauthor{Venkatesh S. Kadandale$^\ast$}{venkatesh.kadandale@upf.edu}{1}
\addauthor{Gloria Haro}{gloria.haro@upf.edu}{1}

\addinstitution{
 Universitat Pompeu Fabra\\
 Barcelona, Spain
}

\runninghead{Montesinos, Kadandale, Haro}{A cappella -- AV Singing Voice Separation}

\def\eg{\emph{e.g}\bmvaOneDot}

\def\ie{\emph{i.e}\bmvaOneDot}
\def\etal{\emph{et al}\bmvaOneDot}

\begin{document}

\maketitle

\begin{abstract}

The task of isolating a target singing voice in music videos has useful applications.
In this work, we explore the single-channel singing voice separation problem from a multimodal perspective, by jointly learning from audio and visual modalities. To do so, we present \textit{Acappella}, a dataset spanning around 46 hours of \textit{a cappella} solo singing videos sourced from YouTube. We also propose an audio-visual convolutional network based on graphs which achieves state-of-the-art singing voice separation results on our dataset and compare it against its audio-only counterpart, U-Net, 
and a state-of-the-art audio-visual speech separation model.
We evaluate the models in the following challenging setups: i) presence of overlapping voices in the audio mixtures, ii) the target voice set to lower volume levels in the mix, and iii) combination of i) and ii). The third one being the most challenging evaluation setup. We demonstrate that our model outperforms the baseline models in the singing voice separation task in the most challenging evaluation setup.
The code, the pre-trained models, and the dataset are publicly available at \url{https://ipcv.github.io/Acappella/}
\end{abstract}

\section{Introduction}

Voices form an integral part of our daily lives. In the form of speech, human voice serves as an effective means of communication. The same voice, when vocalised in sustained tonality and/or rhythm, turns into something musical: the singing voice. The singing voice has become a vital element in the music industry today. Apart from its usage as lead singing voice in songs, it is also found in other diverse forms like rap music, opera singing, solfège, scatting, humming, backing vocals and beatboxing to name a few. \textit{A cappella} refers to a musical arrangement with single or multiple singing voices without any instrumental accompaniment. We are interested in isolating the target voices of interest in multi-voice \textit{a cappella} videos, and in general, in music videos with singing faces.

Singing voice separation has been largely explored in the context of separating voice from the instrumental accompaniment. 
\begin{comment}
The timbral characteristics of singing voice are clearly different from that of the accompanying musical instruments. 
\end{comment}
The audio-only models developed for separating the singing voice from the instrumental accompaniment (\eg \cite{takahashi2018mmdenselstm, samuel2020meta, li2021sams}) largely benefit from the differences in the timbral characteristics between the singing voice and the accompaniment. However, such models do not perform well in the case of separating a particular voice from a mixture of overlapping voices or when the volume of the desired target voice is low. In fact, a very similar problem appears in speech separation when there are overlapping speech segments from different sources in a speech mixture. The audio-visual speech separation methods that leverage the visual information to isolate the desired target speech have been shown to outperform their audio-only counterparts \cite{ ephrat2018looking, gao2021visualvoice, nguyen2020deep, wu2019time}. For an extensive review of audio-visual speech separation works, see \cite{michelsanti2020overview}. Likewise, we are interested in improving upon the audio-only singing voice separation method by incorporating the visual information. We show that using the visual features is particularly advantageous in the singing voice separation task, especially in the aforementioned challenging cases: multi-voice mixtures and mixtures with low volume target singing voice.

A system capable of isolating the target voice of interest in an audio mix has many applications. Such a system could be helpful in evaluating individual singing voices in multi-voice audio mixtures. It can also be useful for automatic karaoke generation, music unmixing and remixing, lyrics and pitch transcription, pitch correction and melodic analysis. 

While there are different audio-visual benchmark datasets for speech separation (reviewed in \cite{michelsanti2020overview}), to the best of our knowledge, to date there is no public dataset available for audio-visual singing voice. One of the contributions of the paper is a new dataset with videos of solo performances of people singing \textit{a cappella}.  
This dataset can be used to train audio-visual networks for singing voice separation or for style/voice conversion. 

We also propose a new audio-visual network for singing voice separation. It is based on a U-Net that processes a complex spectrogram and it is conditioned by the motion features extracted by a spatio-temporal graph convolutional network  that receives a sequence of face landmarks. Although there are recent works that use graph neural networks with face landmarks for face identification \cite{papadopoulos2021face} and emotion recognition \cite{ngoc2020facial}, or with skeletons for separating musical instruments \cite{gan2020music}, to our knowledge, we are the first ones to use face landmarks processed with a graph neural network for audio-visual source separation in the speech/singing context.
The U-Net architecture has been extensively used both in audio-only source separation methods \cite{meseguer2019conditioned, jansson2017singing, kadandale2020multi, stoller2018wave} as well as in its audio-visual counterpart \cite{gao2019co, owens2018audio, zhao2018soundofpixels, zhao2019soundofmotions, xu2019recursive, zhu2020visually}. We can also 
find works on source separation that condition the U-Net on prior information such as  the presence of certain types of musical instruments \cite{slizovskaia2019end, slizovskaia2021conditioned}, phoneme activation for singing voice separation \cite{meseguer2020content} or the fundamental frequency  contour of each type of voice sources in choir ensembles \cite{petermann2020deep}.

In summary, our contributions are three-fold: i) a new dataset of solo singers performing  with no accompaniment, 
ii) a new audio-visual deep neural network for singing voice separation that uses a spatio-temporal graph convolutional network to extract motion features from face landmarks, and
iii) an ablation of four 
different possibilities for the visual network in the audio-visual architecture and two different training settings. 
Both the dataset and  our model are, to the best of our knowledge, the first ones presented in the literature for audio-visual singing voice separation with publicly available code and data for reproducibility. 

\section{Related work}
In the audio-visual speech separation works, there are multiple ways in which the visual features are extracted, depending on the front-end representation of the visual information. Many of such works \cite{nguyen2020deep, wu2019time, afouras2018conversation,li2020deep, gabbay2018visual} operate directly on the mouth region of the video input to extract the lip motion features. In \cite{morrone2019face}, the motion vectors of face landmarks are used as input to an LSTM-based network. On the other hand, \cite{ephrat2018looking} makes use of face embeddings \cite{cole2017synthesizing} extracted on the input video frames containing the whole face. These face embeddings are invariant to illumination, pose, and facial expression. The authors show that, apart from the region around the mouth, the facial parts like eyes and cheeks also contribute to the speech separation performance. A very recent work \cite{gao2021visualvoice} leverages not only the lip motion features but also the facial appearance of the speaker since it is related to certain speech attributes. Their network is trained in a multi-task fashion that jointly learns audio-visual speech separation and cross-modal face-voice embeddings that assist in establishing face-voice mappings.
In \cite{chung2020facefilter}, a single face image of the target speaker is used to condition an audio-visual source separation model based on facial appearance. The correlation of voice traits and facial attributes has also proven useful in speaker identification \cite{kim2018learning} and image generation \cite{oh2019speech2face} tasks.  Further, \cite{fernandez2017towards} points out that facial expressions are helpful in the visual speech recognition task. 

In a concurrent work, Li \cite{li2020multi} explored the specific task of audio-visual singing voice separation. Li's audio-visual singing voice separation method particularly outperformed the audio-only baseline methods when the input sample contained backing vocals in addition to the target voice. 
Our work is along the similar lines but, in addition, we analyse the effect of volume of the target voice on the source separation quality. Further, our approach also differs from Li's work in terms of the choice of baseline models, the proposed model architecture, the experimental setup and the dataset.

\section{The Dataset}

In order to exploit the visual information in the singing voice separation problem, we gathered a new dataset of people singing \textit{a cappella}{, \ie with no music accompaniment}.
The dataset, named \textit{Acappella}, comprises around 46 hours of \textit{a cappella} solo singing videos (\ie a single singer per video) sourced from YouTube, sampled across different singers and languages. It covers four language categories: English, Spanish, Hindi and others. 

The samples in our dataset are defined based on the timestamps corresponding to the segments of interest in each of the videos. These timestamps are provided in the dataset. They have been manually selected to exclude parts of the videos that do not satisfy any of the following characteristics: single frontal face view without occlusions, minimal background noise, no beatboxing, no snapping fingers, songs with lyrics.
\begin{comment}
(\eg we avoid humming and yodelling).
\end{comment}

\begin{figure}[ht]
\centering
    {{\includegraphics[width=5cm]{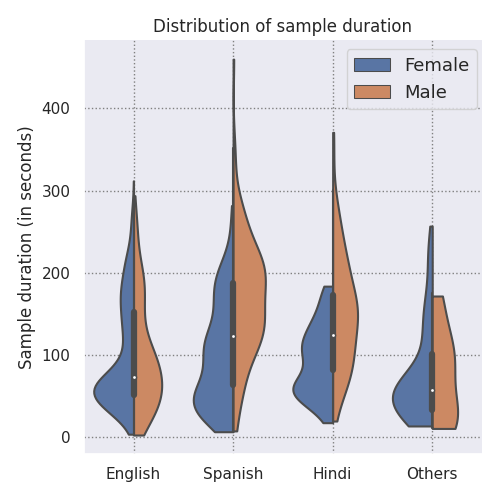} }}%
    \qquad
    \hspace{-2.5em}
    {{\includegraphics[width=5cm]{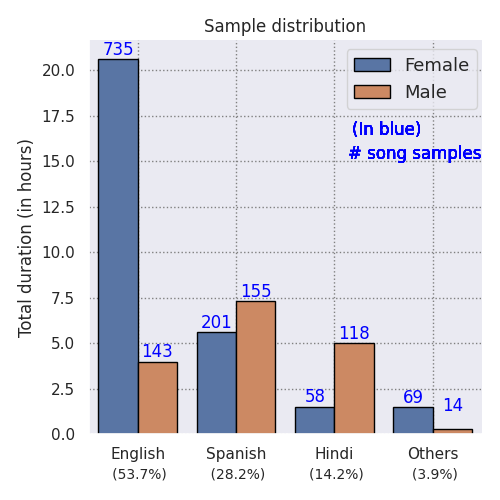} }}%
    \caption{\textit{Acappella} dataset statistics.}%
    \label{fig:dataset}%
\end{figure}

Along with the dataset, we provide the splits for training set, validation set and test set. The training set makes up around 80\% of the total dataset. Around 7\% of the dataset forms the validation set which is used during the training to save the best checkpoint. The test set is divided into the following subsets: seen-heard and {unseen-unheard}. The former consists of samples from known singers, \ie singers {present} in the training set but singing different songs. The latter contains singers who are not a part of the training set. The {unseen-unheard} test subset also contains samples from languages not {heard} in the training set. It 
presents an approximately uniform distribution of samples across language categories and gender. 
Extended statistics of the complete dataset are shown in Figure \ref{fig:dataset}.

Li \cite{li2020multi} created a similar dataset. It comprises of 491 solo singing voice YouTube videos and 65 recorded ones, which overall sum up to 12 hours. To our knowledge, the dataset presented in this paper is the biggest dataset of audio-visual solo singing voice and, at present, the only one which is public.

We also wanted to test our models to separate voices in multi-voice  videos where multiple singing faces are put together in a single view. Since such videos do not provide us with the individual voices for each face, it is not possible to quantitatively evaluate our models on them. Hence, we assembled  a multi-voice video ourselves. The mixture contains six  voices sung by the same person. The lead voice content is in English and Zulu, there is a voice emulating a flute, and the rest pair up and sing in unison most of the time in Zulu. 
Background accompaniment music is also included in this mixture to add to the complexity.

\section{Singing voice separation model}
\label{sec:architecture}
Our model architecture comprises of a multimodal convolutional neural network which takes in a video and its corresponding mixture audio waveform and returns a complex mask. 
The waveform is mapped into the time-frequency domain using a short-time Fourier transform (STFT). The estimated mask allows to recover the separated voice of the target singer by computing the complex product between the mask and the spectrogram. 

Our network is designed to receive only the visual information of the target singer to isolate, mainly for two reasons: i) it allows to reduce and bound the memory required for training, and ii) it broadens the applicability of the model since it only needs to be shown the face of the target singing voice with no additional visual information related to the other sources. This way, the model can address mixtures of singing voice with accompaniments of different nature: musical instruments, backing vocals, other lead voices, beatboxing, snapping fingers, ambient sounds, or even different types of noise.

\begin{figure}[ht]
\centering
\includegraphics[width=\linewidth]{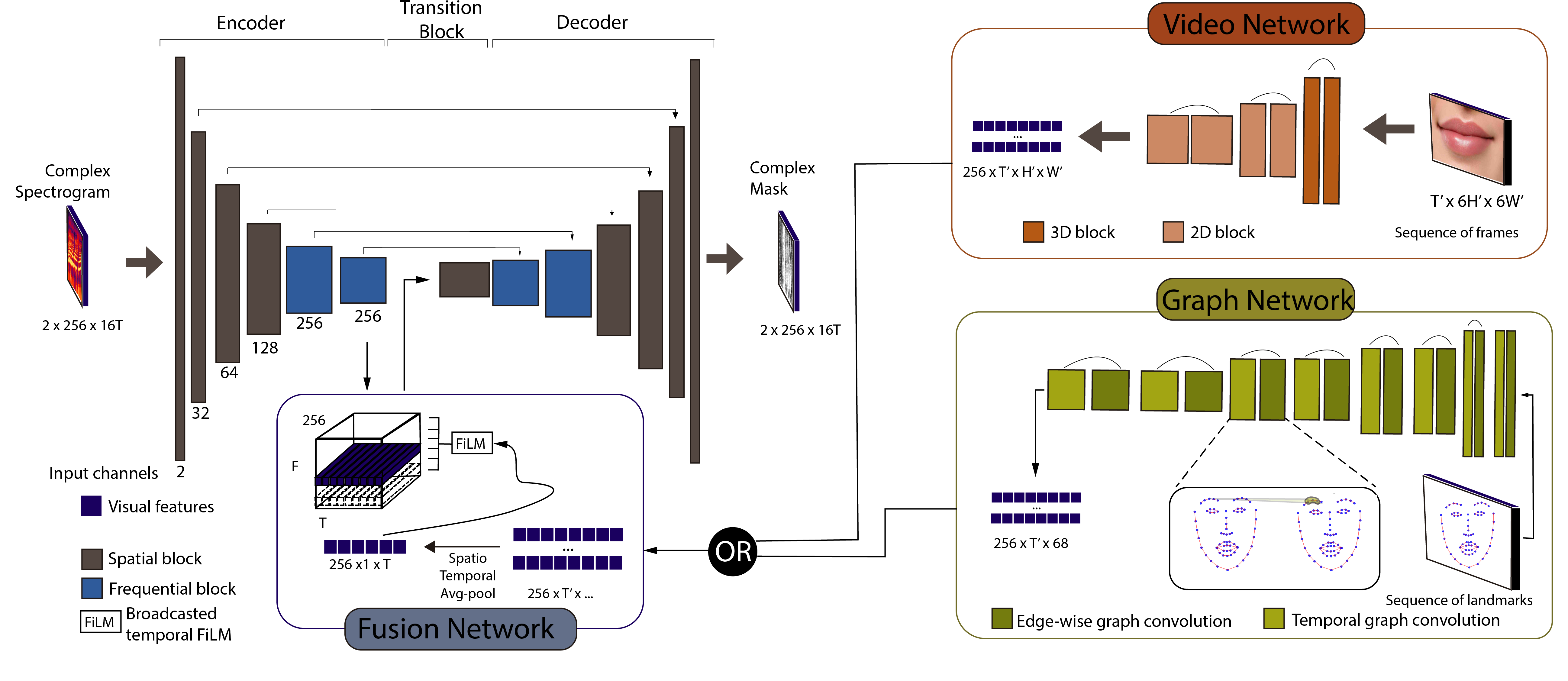}
\caption{Y-Net model scheme. The system works with chunks of $4n$ seconds, where $n \in N$. The audio network takes as input a $256\times 16 T n$  complex spectrogram and returns a complex mask. The visual network in case of Y-Net-m and Y-Net-mr, is the video network (in red), which takes as input a set of 100$n$ frames cropped around the mouth of the target singer. In case of Y-Net-g and Y-Net-gr, the visual network is the graph network (in green) which takes as input a sequence of 68$n$ landmarks of the face of the target singer. The visual features are fused with the audio network's latent space through a FiLM layer (we use $T=16$). The FiLM broadcasts the $256\times 1 \times T$ visual features into the $256\times16\times T$ audio ones. The spatial blocks of the U-Net downsample in both, the frequency and the temporal dimension, while the frequential block downsamples along the frequency dimension only.}
  \label{fig:model}
\end{figure}

The architecture is a two-stream convolutional  neural  network for processing video and audio. It is denoted as Y-Net and illustrated in Figure \ref{fig:model}. The audio network consists of a 6-blocks U-Net which predicts a two-channel tensor. The U-Net \cite{ronneberger2015u} is an encoder-decoder architecture with skip connections in between which allows to preserve the spatial structure while increasing the receptive field through blocks.
We have experimented with two different number of blocks in the U-Net 
(see a comparison in Table \ref{tab:ablation}) to ensure the best performance without overfitting. The original U-Net design doubles the amount of channels each block while it downsamples the spatio-temporal resolution by two. In our U-Net with six blocks, we keep both the temporal resolution and the amount of channels, fixed, in the last blocks (\ie the features are downsampled only along the frequency domain).
The rationale behind this is not to lose much temporal resolution so that the features coming from the visual modality can be aligned to the audio ones and condition on those.
We fix the temporal resolution of the U-Net bottleneck to 16 frames; this ensures that there are no out-of-synchronisation issues between both modalities and at the same time ensuring a fine enough temporal resolution for the separation task.
On the other hand, a recent work  \cite{lee2021looking} applies a synchronisation module between video and audio modalities but they  deal with strong miss-alignments (up to 0.36s) which is not the case in our videos.

For the video network, we experiment with four different options:

1) {\bf Y-Net-g:} This network extracts motion features from a sequence of aligned face landmarks (more details in Section \ref{sec:pre-proc}). The reasons for using landmarks instead of raw frames are:  i) a sequence of face landmarks contains motion information of the face, ii) appearance information and video background are removed, making the system less prone to overfitting, and iii) the computation cost is much less compared to processing of the video frames.  To make profit of the landmarks, we use a variation of the spatio-temporal graph convolutional neural network from \cite{yan2018spatial}, denoted as ST-GCN. Graph CNNs are a generalisation of traditional convolutions. Akin to the traditional convolutions, given a root node, ST-GCN works on a neighbour set of nodes as shown in Figure \ref{fig:model}.  We treat the face  landmarks as undirected graphs, where nodes encode the cartesian position of each landmark on the image. 

2) {\bf Y-Net-m:} A sequence of video frames cropped around the mouth (more details in Section \ref{sec:pre-proc}) are fed to a 3-block 3D-ResNet-like network, where the first block is  3D convolutional  and the last two blocks are 2D convolutional. The 3D convolutional block processes motion information. This design turns into a network with 3M parameters (M stands for million). In contrast, a  traditional 3D-ResNet18 has 33.4M and the 2D-ResNet18 has 11.4M. This way, the visual network keeps the capacity to model spatio-temporal information, as suggested in Tran \etal \cite{tran2018closer}, while having a contained amount of parameters not to overfit. 

3) {\bf Y-Net-e:} We consider the visual network used in Ephrat \etal \cite{ephrat2018looking}. The input to this visual network are the face embeddings extracted from the video frames cropped around the face, just like in  \cite{ephrat2018looking}. 
The visual network comprises of six 1D dilated convolutional blocks.

We also experiment with the following additional configuration for the video network:

4) {\bf Y-Net-f:} While the Y-Net-m ingests a sequence of video frames cropped around the mouth, Y-Net-f takes in a sequence of video frames cropped around the entire face. More details in Section \ref{sec:pre-proc}.

The visual features are fused with the audio networks' latent features via FiLM conditioning \cite{dumoulin2018feature}. Note that since both the audio and visual features are processed with convolutions, the time-frequency and spatio-temporal structures are kept, allowing to fuse them after an alignment in the temporal dimension. We apply a spatio-temporal average pooling to the video features to get the same number of features in  the temporal dimension as the audio ones. At inference time, the model can work with chunks larger than 4s, only limited by the available memory, enabling a fast processing in contrast to processing chunks of 4s and concatenating the resulting masks which could introduce artifacts.

\subsection{Pre-processing} \label{sec:pre-proc}
\textit{Video processing.} Videos are resampled to 25 fps to maintain uniform sampling rate across all the samples. We pre-processed the video stream of the target singer using a face detector\footnote{https://github.com/DinoMan/face-processor} to extract 68 face landmarks, cropping around the face and aligning the face  along all the frames in the video.
In case of Y-Net-m, each frame is cropped around the mouth region and then resized to $96 \times 96$. Whereas, for Y-Net-f, each frame is cropped around the full face region and then resized to $128 \times 96$.Then, we feed the visual network with a sequence of 100 RGB frames, corresponding to 4s of video. 
In case of Y-Net-g, we feed the spatio-temporal graph network with the aligned sequence of face landmarks.

\noindent \textit{Audio processing.} The audio signal is resampled to 16384 Hz. We consider a 4s-audio excerpt and compute its STFT using a  Hanning   window of size 1022 and a hop length of 256 (as in \cite{zhao2018soundofpixels, gao2019co}) which leads to a 512$\times$256 spectrogram. This specific shape is useful to achieve a perfect alignment between the downconvolutional and the upconvolutional blocks of the U-Net, which are connected through the skip connections. For computational efficiency, we downsample the spectrogram in the frequency dimension and use a 256$\times$256 spectrogram. 

\subsection{Training strategy, training target and loss}
We train the networks in a self-supervised way by generating the audio mixtures artificially. Given a set of N waveforms, ${s_1, ..., s_N}$, we generate an artificial mixture by taking the average, \ie  $s_m=\frac{1}{N}\sum s_i$. This way we can ensure the resulting mixture is bounded between -1 and 1. The network is trained to optimise an $L_2$ loss on bounded complex ratio masks \cite{williamson2015complex}.

Let $S_i(f,t)$  be the STFT of a generic waveform $s_i$. Note that $S_i(f,t)$ is a complex matrix. We define the ideal complex ratio mask  as follows:
$M(f,t)=\frac{S_i(f,t)}{\sum S_i(f,t)}.$

Since the mask $M$ 
is not bounded, we apply a hyperbolic tangent on the real and imaginary parts, $M^r$ and $M^i$, respectively, to obtain a bounded complex mask: 
\begin{equation} \label{bounded_mask}
M_b(f,t) = M_b(f,t)^r + M_b(f,t)^i i =  \tanh({M^r(f,t)})+\tanh({M^i(f,t)}) \, i.
\end{equation}

Let $\hat{M}_b$  be the bounded mask estimated by the network. The loss function is defined as:
\begin{equation}
\mathcal{L}={\|G^\frac{1}{2}\odot(\hat{M}_b^r-M_b^r)\|}_2^2+{\|G^\frac{1}{2}\odot(\hat{M}_b^i-M_b^i)\|}_2^2,
\end{equation}
where $\odot$ denotes the element-wise product and $G$ is 
a gradient penalty so that the points of the mixture spectrogram $S_m$ with higher energy contribute more to the loss, it is defined as:
\begin{equation} \label{gradient_penalty}
G(f,t) = \max(\min(\log (1+|S_m(f,t)|),10),10^{-3}).
\end{equation}

A very common problem with multimodal networks is how to force the model to pay attention to one modality when the  task is easy to solve from the other modality alone. 
When the patterns of each sound source are clearly different, the source separation is easier from the audio modality. Thus, we artificially create mixtures with different types of accompaniments, including human voices. 
Since we only need the face of the target singer, we  mix samples from \textit{Acappella} together with samples from Audio Set \cite{gemmeke2017audio}. Audio Set is an in-the-wild large-scale dataset of audio events across more than 600 categories. We gathered the categories related to the human voice and some typical accompaniments. These categories are: acappella,  background music,  beatboxing,  choir,  drum, lullaby, rapping, theremin, whistling and  yodelling. We also include pop and rock music accompaniment from MUSDB18 dataset \cite{musdb18}. While creating artificial mixtures, we ensure that all the samples from \textit{Acappella} are used in each epoch. Those are mixed with a random sample from Audio Set or MUSDB18. We uniformly sample from all the accompaniment set categories. Including Audio Set in the training strategy increases the robustness of the model and addresses overfitting. 

We consider different variants of our model: Y-Net-g, Y-Net-m, Y-Net-e and Y-Net-f. Note that these models, when referred to, without any additional suffix, indicate that they have been trained with mixtures that only contain one lead singing voice which is sourced from \textit{Acappella} dataset and mixed with an accompaniment sample sourced from Audio Set or MUSDB18.
On the other hand, we further append the suffix `r' to the model name to indicate that it has been trained with mixtures in which, 50\% of the time, the mixture contains an additional lead singing voice sourced from \textit{Acappella} dataset. In this work, the experiments with model Y-Net-f are limited to its respective `r' variant, Y-Net-fr, only.

\section{Experiments}
We conduct a set of experiments comparing the different Y-Net versions against their audio-only counterpart, the U-Net (\ie our Y-Net without the visual network), and a state-of-the-art model for speech separation, the model of Ephrat \etal  \cite{ephrat2018looking}, that we denote as LLCP\footnote{we use an existing code available at https://github.com/vitrioil/Speech-Separation}(more details in the supplementary material). Results are expressed in terms of Signal-to-Distortion Ratio (SDR) and  Signal-to-Interference Ratio  (SIR), both defined in \cite{sdr}.

We are interested in analysing the role of  different types of  visual information in  different kind of mixtures. For that, 
we evaluate the models in two different setups: mixing a single singing voice with accompaniment (one lead voice setup) and mixing two singing voices with accompaniment (two lead voices setup). Note that the singing voice(s) in both these setups are always sourced from \textit{Acappella} dataset.
Experiments are conducted both for seen-heard and unseen-unheard singers in heard languages and unseen-unheard singers in unheard languages (\ie new languages) to check how  the different networks generalise.

 \begin{table}[ht]

 \footnotesize
 \centering

        \begin{tabular}{c|cc|cc}
        Models  & \multicolumn{2}{c|}{4-blocks U-Net} & \multicolumn{2}{c}{6-blocks U-Net} \\ \cline{2-5}
                & SDR          & SIR          & SDR          & SIR          \\\hline \hline
        U-Net   &      -1.92    &      12.16  &     -1.97   &      12.64        \\
        Y-Net-e &      -1.50    &      12.50  &     --       &      --        \\
        Y-Net-m   &     {\bf 2.49}    &      14.04  &     {\bf 2.91}    &      {\bf 15.71}        \\
        Y-Net-g  &       1.85    &      {\bf 14.42}  &     2.07    &     15.49  
        \\\hline
        Y-Net-fr  &      --    &      --  &     4.54   &    15.39 \\
        Y-Net-mr  &      3.38    &      13.81  &     5.03    &    15.80 \\
        Y-Net-gr  &      {\bf 4.71}    &      {\bf 15.67}  &     {\bf 6.41}    &    {\bf 17.38}
        \end{tabular}
       % }
        \caption{Ablation study on the unseen-unheard test set in the two lead voices setup.}
        \label{tab:ablation}
\end{table}

In Table \ref{tab:ablation}, we show an ablation study of our model in the unseen-unheard test set in the two lead voices  setup. We analyse four different aspects: i) the number of blocks in the U-Net, ii) audio-only versus audio-visual models, iii) the type of visual network, and iv) the training setting. First thing to notice is that the audio-only model, U-Net, performs much worse than the audio-visual ones and is the only model that does not get benefited from an increase of the number of blocks, since two lead voices are harder to separate from audio alone (actually, in the one lead voice setup U-Net does improve with more blocks). Thus, an increase in the U-Net blocks in the Y-Net models implies a gain in performance 
since the visual information is added to the network, which is indeed a crucial information to get a proper separation in the two lead voices setup. Second, both Y-Net-m and Y-Net-g perform better than Y-Net-e; 
from that, we hypothesise that visual embeddings do not sufficiently encode motion information. This follows the observations of \cite{cole2017synthesizing}, which explains that visual embeddings ignore factors of variation related to aspects such as lighting, pose and expression (the latter being more related to the face motion).
The Table \ref{tab:ablation} results also show how a boost in performance can be achieved if we train our models with mixtures in which 50\% of the time two lead voices are present (`r' variants). This boost is particularly high (+2.86 dB and +4.34 dB in SDR, in 4-blocks and 6-blocks respectively) in the graph-based model, Y-Net-g, compared to the video-based model, Y-Net-m (+0.89 dB and +2.12 dB). Finally, we explore further (with the `r' variants), which among the full face and the cropped mouth, works best as input to the video network; for that we include the results of the Y-Net-fr variant. Y-Net-fr relies on the same network as Y-Net-mr but the input to the video network are crops of the frames containing the full face {rather than the} lips region {alone} as in Y-Net-mr. Both the SDR and SIR  values indicate that a better separation with the video network is achieved {by limiting the visual information only to the lips region}
{As it can be observed,} the best model is Y-Net-gr with 6-blocks U-Net. 
From here on, all our model variants use a 6-blocks U-Net.

\begin{table}[ht]
\resizebox{\linewidth}{!}{%
\begin{tabular}{l|cccc|cccc|c}
Model& \multicolumn{4}{c}{Seen-Heard}                                                            & \multicolumn{4}{|c|}{Unseen-Unheard}                                                          & Multi-voice \\\cline{2-10}
                       & English & Spanish & Hindi & Others & English & Spanish & Hindi & New Languages & English + Zulu                             \\\hline\hline
U-Net                & -1.89                    & -2.25                    & -2.72                  & -1.42                           & -1.86                    & -2.34                    & -1.92                  & -2.15                             &       5.18                      \\
LLCP       \gh{}{\cite{ephrat2018looking}}            & -0.55                    & -0.57                    & -1.08                  & -0.58                           & -0.9                     & -1.18                    & -0.73                  & -1.27                             &            5.63                  \\
Y-Net-m               & \bf{ 4.17}                   & \bf{3.60 }                     & \bf{3.50  }                  &  \bf{3.19   }                         & \bf{3.28}                     &\bf{ 3.33  }                 & \bf{ 2.11 }                 & \bf{ 2.31}                             &               \bf{7.24}               \\
Y-Net-g               & 2.98                     & 2.30                      & 1.79                   & 2.18                            & 2.47                     & 2.74                     & 1.53                   & 1.74                              &                 6.72             \\ \hline
Y-Net-mr              & 7.78                     & 5.42                     & 5.32                   & 5.82                            & 5.33                      & 5.14                     & 4.35                  & 4.07                              &             6.51                 \\
Y-Net-gr             & {\bf 8.61}                     & {\bf 6.62}                     & {\bf 5.91}                   & {\bf 7.45   }                         & {\bf 6.73 }                    & {\bf  6.72 }                    & {\bf 5.76  }                 & {\bf 5.27}                              &              \bf{ 7.21}             
\end{tabular}
}
\caption{SDR results in the two lead voices setup for different methods across languages, both in seen-heard and unseen-unheard test sets. SDR results also for the multi-voice case.}
\label{tab:comparison}
\end{table}

Table \ref{tab:comparison} presents a comparison of our best models, Y-Net  with video network or with graph network, with respect to the U-Net and LLCP models in both the seen-heard and unseen-unheard test sets in the two lead voices setup,  as well as in the multi-voice recording  with ground truth sources (singer not present in the training set). The SDR metrics are shown for different languages. We can observe that across models, the general tendency is that  the performance increases for the languages more represented in the training set. 
Again,  U-Net performs the worst. LLCP outperforms U-Net but not our models. Models that have been trained with 50\% of the samples containing two lead voices (`r' variants) have a boost in performance. Overall, it seems that the graph-based network can better exploit the motion information if the network is trained with the proper mixtures (two lead voices). The boost in performance with the Y-Net-gr is considerably and consistently higher than the boost with the Y-Net-mr, with an average boost of +1.89 dB for Y-Net-m  and +3.98 dB for Y-Net-g.

\begin{figure}[ht]
\centering
  \includegraphics[width=0.75\linewidth]{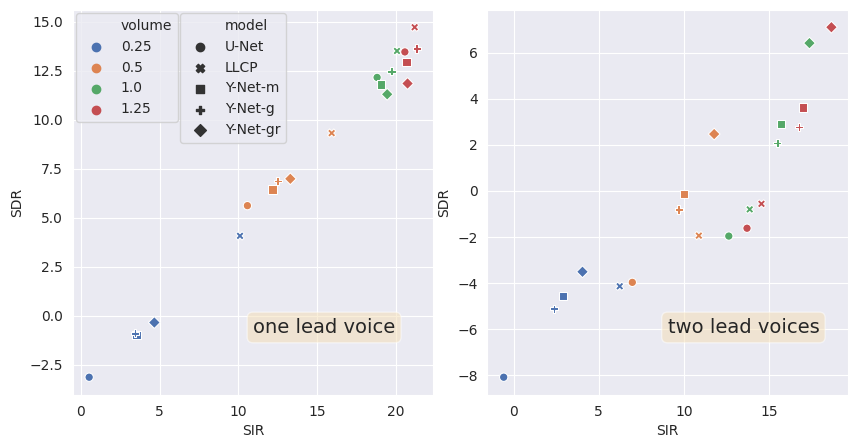} 
  \caption{Results in the unseen-unheard test set: left, one lead voice setup; right, two lead voices setup. Different symbols are assigned to the different models and different colours to the different volume levels of the target voice.}
  \label{fig:volume}
\end{figure}

In order to evaluate how sensitive are the different models to the volume of the target singer, we use different volume levels in the singing voice, so that experiments range from predominant singing voice to non-dominant one. To do so,  
each source $s_i$ in the mixture 
is normalised by its root mean square value and then the singing voice is further  multiplied by a factor
 $\alpha$, where $\alpha \in \{0.25, 0.5, 1, 1.25\}$. 
Lastly, we rescale {all} the sources {with the same value} to ensure they are bounded between -1 and 1 while respecting the relative preset volumes {(we divide by the maximum of absolute values of all sources)}. Figure \ref{fig:volume} shows metrics for different volume levels and different methods in the unseen-unheard test set in two different setups: one lead voice (left) and two lead voices (right). 
For the one lead voice setup, LLCP performs the best in all volume levels. The second best for volume level factors of 1.25 and 1 is Y-Net-g, while for lower volume  factors, 0.5 and 0.25, the second best is Y-Net-gr. Y-Net-gr exploits the motion information more than Y-Net-g and Y-Net-m since it has been trained with mixtures containing two lead voices, where motion is a key factor. This result shows that motion is also important in the case of one singing voice with a low volume, where the audio information alone is not enough to perform a good separation. Actually, we can observe how the 
In case of two lead voices, the Y-Net-gr is the best model for all volume levels (except for SIR in the lowest volume case, where it is the second best). The rest of our models are better than LLCP for volume levels of 1.25 and 1 (both in SDR and SIR) and better than LLCP in terms of SDR for volume level of 0.5. Overall, we can conclude that LLCP is a good choice for the one lead voice case and the Y-Net-gr is the best model for two lead voices, where the motion features become crucial to get good separation results.

\begin{table}[htb]
\footnotesize
\centering
\begin{tabular}{r|cc|cc|cc}
Y-Net-gr  & \multicolumn{2}{c}{One lead voice in test}   &  \multicolumn{2}{|c}{Two lead voices in test} &  \multicolumn{2}{|c}{Average}\\
 Remix \%     &  SDR & SIR &  SDR & SIR &  SDR & SIR\\\hline\hline
0 \%               & {\bf 12.47} & {\bf 19.71} & 2.07  & 15.49 & 7.27  & 17.6 \\
50 \%              & 11.29 & 19.43 & 6.41  & {\bf 17.38} & {\bf 8.85}  & {\bf 18.41}\\
75 \%              & 11.08 & 18.78 & {\bf 6.42}  & 17.09 & 8.75  & 17.93\\
100 \%             & 9.98 & 17.73 & 6.40  & 16.68 & 8.19  & 17.21\\                  
\end{tabular}
\caption{Ablation study on the percentage of mixtures containing two lead voices in the training of the Y-Net-gr model (note that 0\% corresponds to the Y-Net-g model). Results on the unseen-unheard test set.}
\label{tab:remix_percentage}
\end{table}

Finally, Table \ref{tab:remix_percentage} shows an ablation on the percentage of mixtures with two lead voices in the training set in the case of our best model,  Y-Net-gr. 
The table shows the performance of these models both in the one lead voice and two lead voices setup in the unseen-unheard test set. 
The increase in the percentage of two lead voice mixtures while training degrades the test results in the one lead voice setup when the volume of the singing voice is reasonable ($\alpha=1$, as it is the case both in the training and test sets of this ablation). However, as seen in Figure \ref{fig:volume} (left), when the volume of the target singing voice is low, training the network with two lead voices helps as well -- note that Y-Net-gr performs better than Y-Net-g also for one lead voice. 
When we evaluate the models with mixtures containing two lead voices, we observe a boost in performance of the different Y-Net-gr models (with different percentages of two lead voice mixtures) with respect to that of the Y-Net-g.
By considering the average results of the one lead and two lead voices setups in the unseen-unheard test set (two rightmost columns in Table \ref{tab:remix_percentage}), we infer that the best model is the one trained with 50\% of the mixtures containing two lead voices. 
Thus, we consider Y-Net-gr model trained with 50\% of mixtures containing two lead voices, as our proposed model as it  achieves a good compromise in both scenarios and also in the case of a target voice with a low volume.

For demos, please visit the project page: \small{\url{https://ipcv.github.io/Acappella/}}.

\section{Conclusions}
This paper explores the singing voice separation problem from a new perspective, by exploiting both the audio and visual information. We introduce a new dataset of video recordings of \textit{a cappella} solo performances. We also propose a new audio-visual singing voice separation model, based on a U-Net conditioned on the motion of the face landmarks of the target singer. Those landmarks are processed with a spatio-temporal graph convolutional network.
Moreover, we present a thorough ablation study of our model, with different variants of the visual network and show how  the performance can be boosted in multi-voice cases by adding mixtures with two lead singing voices in the training set. The experiments show how audio-visual methods improve upon audio-only ones in challenging scenarios when there are multiple overlapping voices or when the target voice has a low volume. The presented model is compared to a state-of-the-art audio-visual speech separation model trained in the new dataset. Our model better exploits the face motion and thus outperforms the baseline models in singing voice separation in the most challenging evaluation setup.

\section*{Acknowledgements}
The authors acknowledge support by MICINN/FEDER UE project, ref.~PGC2018-098625-B-I00; H2020-MSCA-RISE-2017 project, ref.~777826 NoMADS; ReAViPeRo network, ref. RED2018-102511-T; and Spanish Ministry of Economy and Competitiveness under the Mar{\'i}a de Maeztu Units of Excellence Program (MDM-2015-0502) and the Social European Funds. J. F. M. acknowledges support by FPI scholarship PRE2018-083920. V. S. K. has received financial support through “la Caixa” Foundation (ID 100010434), fellowship code: LCF/BQ/DI18/11660064. V.S.K has also received funding from the European Union’s Horizon 2020 research and innovation programme under the Marie Skłodowska-Curie grant agreement No.~713673.
%We also thank NVIDIA Corporation for the donation of GPUs. 
We gratefully acknowledge NVIDIA Corporation for the donation of GPUs used for the experiments.
We thank Emilia Gómez and Olga Slizovskaia for insightful discussions on the subject. %Spanish Ministry of Economy and Competitiveness under the María de Maeztu Units of Excellence Program (MDM-2015-0502) and the Social European Funds.
%This work was funded in part  Spanish Ministry of Economy and Competitiveness under the Mar{\'i}a de Maeztu Units of Excellence Program (MDM-2015-0502) and the Social European Funds; the  MICINN/FEDER UE project with reference PGC2018-098625-B-I00; and the H2020-MSCA-RISE-2017 project with reference 777826 NoMADS. 

\bibliography{egbib}
\end{document}